# The Role of Geometry in Tailoring the Linear and Nonlinear Optical Properties of Semiconductor Quantum Dots

Grigor A. Mantashian[1,2]

1 - Institute of Chemical Physics after A. B. Nalbandyan NAS RA, Yerevan, Armenia;
2 - Russian-Armenian University, Yerevan, Armenia.
grigor.mantashian@rau.am

The paper aims to reveal the relationship between the geometrical features and linear and nonlinear optical properties of InAs quantum dots (QDs). This problem is justified by the extreme variety offered by the recent advances in growth techniques tailored to the attainment of QDs and nanostructures with virtually any shape. To that end, the Finite Element Method in conjunction with the Effective Mass Approximation and Envelope Function Approximation was employed to solve the one-particle eigenproblems in domains with any complex geometries. The paper explores nanoplatelets, spherical QDs, nanocones, nanorods, nanotadpoles, and nanostars. It has been found that there is a clear correlation between the complexity and symmetry of the QDs and their linear and nonlinear absorption spectra for transitions between the electronic ground state and the first three excited states.

## 1. Introduction

Semiconductor nanostructures particular structures with three-dimensional size quantization like quantum dots (QDs) have long been studied and considered important building blocks of nanotechnology. These structures have found their applications in a myriad of fields starting from single-photon sources, and display technologies and ending with flexible biocompatible sensing platforms and biosensing. The cause for such diversity is the endless potential for the optimization of these structures. The size quantization effect depends on various properties of QDs such as material composition, size of the structure, and the shape of the structure. This paper focuses on shape engineering which is perhaps the most flexible method for changing the properties of QDs. By changing the shape of QDs we can change the behavior of the energy spectrum as well as the behavior of the inter-level transition selection rules. These changes in behavior have contributed to a multitude of quantum phenomena from, photoluminescence to single photon sources. This method is further solidified by the fact that nanostructures with any geometrical properties have been attained[1–12] facilitated by advancing growth and synthesis methods such as colloidal synthesis, molecular beam epitaxy, etc. However, despite this experimental development, there is a noticeable disjunction between experimental and theoretical investigation. This is caused by the lack of theoretical tools suitable to flexibly and swiftly design and optimize the properties of QDs. There are two approaches for investigating QDs with complex geometries approximate methods and computational methods. The approximate methods include the geometrical adiabatic approximation methods which can be used to obtain the spectrum and wavefunctions for the oblate and prolate shapes [13–21], and conformal mapping methods which allow to reduce an unsolvable three-dimensional problem into lower dimensions causing a possible loss of information[22–25]. The next is numerical modeling, which by far provides the best flexibility in terms of the shapes of QDs. However, there are many numerical simulation methods with varying degrees of accuracy and computational complexity. By far the most accurate methods are DFT-related methods, with the right implementation they can have a quantitative agreement with experimental data[26–29]. However, these methods have various limitations created by the computational complexity. There are limitations in terms of the size of QDs, if structures are too large, they require either other methods or extremely high computational capacities. A good middle ground is reached by using the finite element method (FEM) alongside the effective mass approximation and envelope function approximation [30]. This ensemble of methods allows the investigation of QDs with various geometrical structures by modifying the solution domain for a one-particle problem. Additionally, by applying various model potentials inside of the Hamiltonian we can investigate different material compositions and band structures such as core/shell, core/shell/shell, type II band alignment[30–35,] etc. Moreover, this method simplifies the consideration of the external electrical, magnetic, and laser fields by modifying the terms inside of the Hamiltonian[36–40].

Intraband optical properties have garnered significant interest due to their implications for novel optoelectronic devices. Studies have elucidated that the size, shape, and material composition of QDs play pivotal roles in dictating their intraband optical behaviors, such as absorption coefficients and energy level transitions[41–46]. For instance, heavily n-doped PbS colloidal QD solid-state films demonstrate a strong intraband absorption, which is as potent as their interband counterparts, showcasing a direct correlation between the QD size and the intraband energy transition span [41]. These transitions exhibit unique temperature dependencies, diverging from traditional interband behaviors and highlighting the nuanced control over optical properties achievable through QD manipulation. Furthermore, the introduction of external stimuli, such as magnetic fields, has been shown to influence intraband optical properties in asymmetric biconvex lens-shaped QDs, suggesting avenues for tunable optoelectronic applications based on intraband transitions[42]. These insights into intraband optical properties not only advance our understanding of quantum confinement effects but also open up new possibilities for the development of infrared photodetectors and light-emitting devices, leveraging the unique intraband transitions offered by QDs.

The primary objective of this study is to explore the intricate relationship between the geometrical features of semiconductor InAs QDs and their intraband linear and nonlinear optical properties. This exploration is grounded in the hypothesis that the shape and symmetry of InAs QDs significantly influence their optical behaviors, including their total intraband absorption coefficients. To that end, the electronic wavefunctions and energies for spherical, cylindrical, and conical QDs have been calculated together with more exotic structures such as nanoplatelets, nanostars, and nanotadpoles.

## 2. Materials and Methods

The material parameters that we are going to use in our calculations for InAs are taken: $m_e^* = 0.023 m_0$ - electron effective mass, $\varepsilon_s = 8.85$ - static dielectric susceptibility, $\varepsilon_\infty = 15.15$ - high-frequency dielectric constant, $E_R = 1.35\,meV$ - electron effective Rydberg energy, $a_{bohr} = 35.03\,nm$ - electron Bohr radius. The volume for all of the structures was kept constant at around $667\,nm^3$, the dimensions for each structure are given in Table 1 at the end of this section. In the article at hand, the computations were performed using the commercial software known as Wolfram Mathematica. Nonetheless, the methodologies and findings apply to alternative software options such as MATLAB, COMSOL Multiphysics, etc. The initial phase in any Finite Element Method (FEM) computation involves establishing the partial differential equation, with the intention in this instance to tackle a three-dimensional single particle Schrödinger equation.

$$-\frac{\hbar^2}{2m^*}\left(\frac{\partial^2}{\partial x^2}+\frac{\partial^2}{\partial y^2}+\frac{\partial^2}{\partial z^2}\right)\psi(x,y,z)+V\psi(x,y,z)=E\psi(x,y,z) \tag{1.1}$$

Here $\psi(x,y,z)$ - is the wavefunction of the single electron, $V$ - is the confinement potential for the electron, and $E$ - is the energy of the electron.

After determining the energy levels and associated wave functions, it's possible to compute both the linear and third-order nonlinear optical absorption coefficients. These coefficients can be obtained using the density matrix method and the perturbation expansion technique, as documented in references[47-49]. Within the framework of a two-level system, the expressions for both linear and nonlinear third-order optical absorption coefficients in a quantum dot are presented as follows:

$$\alpha^{(1)}(\omega)=\frac{\sigma e^2}{\hbar}\sqrt{\frac{\mu}{\varepsilon}}|M_{fi}|^2\frac{\hbar\omega\hbar\Gamma_{fi}}{\left(E_{fi}-\hbar\omega\right)^2+\left(\hbar\Gamma_{fi}\right)^2} \tag{1.2}$$

$$\alpha^{(3)}(\omega)=\frac{2\sigma e^4}{\hbar}\sqrt{\frac{\mu}{\varepsilon}}\left(\frac{I}{\varepsilon_0 n_r c}\right)|M_{fi}|^4\frac{\hbar\omega\hbar\Gamma_{fi}}{\left(\left(E_{fi}-\hbar\omega\right)^2+\left(\hbar\Gamma_{fi}\right)^2\right)^2}$$
$$\times\left(1-\frac{|M_{ff}-M_{ii}|^2}{4|M_{fi}|^2}\times\frac{3E_{fi}^2-4\hbar\omega E_{fi}+(\hbar\omega)^2-\hbar(\hbar\Gamma_{fi})^2}{E_{fi}^2+(\hbar\Gamma)^2}\right) \tag{1.3}$$

where $e$ - electrons charge; $\sigma$ - charge concentration divided by system volume $\sigma=\frac{n}{V}=\frac{3n}{4\pi a^2 c}$; $\mu$ - system magnetic permeability; $\varepsilon$ - the dielectric constant; $n_r$ - relative refractive index; $\hbar\omega$ - the energy of the incident photon; $E_{fi}=E_f-E_i$ - the difference between the energies of the initial and final states; $\Gamma_{fi}=1/\tau_{fi}$ - it is an off-diagonal matrix element known as the relaxation rate of the final and initial states and is relaxation time; $M_{fi}=\langle\psi_i|z|\psi_f\rangle$ - transition matrix element; $I$ - the intensity of the incident electromagnetic radiation. The total absorption coefficient $\alpha(\omega,I)$ will be equal to the sum of the linear and nonlinear optical absorption coefficients

$$\alpha(\omega,I)=\alpha^{(1)}(\omega)+\alpha^{(3)}(\omega,I) \tag{1.4}$$

**Table 1.** Geometrical parameters of the structures investigated in the paper.

| Nanostructures | Dimensions |
|---|---|
| nanoplatelet | $h_x = h_y = 35nm$, $h_z = 8.75nm$ |
| spherical QD | $r \approx 13.74nm$ |
| nanorod | $r \approx 8.75nm$, $h \approx 45.15nm$ |
| nanocone | $r \approx 14nm$, $h \approx 53.375nm$ |
| nanotadpole | $r_{head} \approx 12.25nm$, $r_{tail} \approx 6.125nm$, $h \approx 26.25nm$ |
| nanostar | $r_{core} \approx 11.375nm$, $r_{cone} \approx 8.75nm$, $h_{cone} \approx 24.5nm$ |

## 3. Results and Discussion

In the current section, the results for the total intraband absorption coefficient, which depends on the incident light energy, are presented. This is accompanied by an analysis of the first four electron states participating in the aforementioned absorption events. These findings are showcased for all structures enumerated in Table 1 with corresponding parameters, highlighting the impact that shape engineering can have on the properties of zero-dimensional nanostructures.

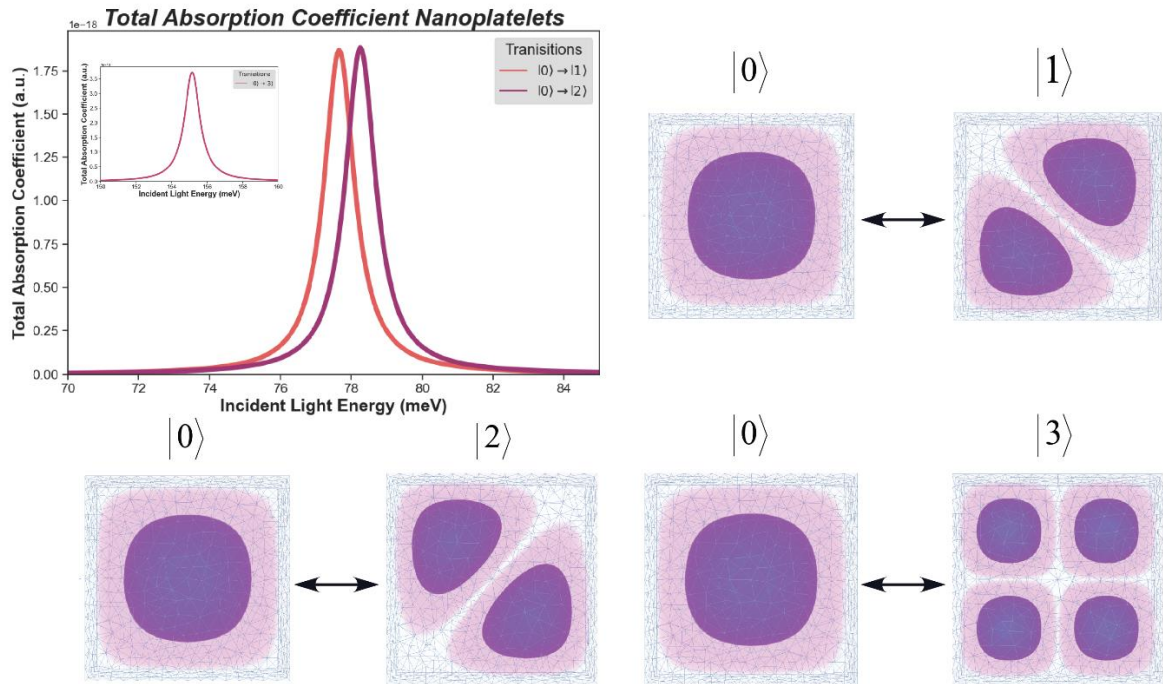

**Figure 1.** The dependence of the total intraband absorption coefficient on the incident light energy is presented in the left upper corner for the following transitions: $|0\rangle \rightarrow |1\rangle$, $|0\rangle \rightarrow |2\rangle$ for an electron confined in an InAs nanoplatelet. In the inset figure the total absorption for the $|0\rangle \rightarrow |3\rangle$ transition is given. In the plots surrounding the main figure, the probability densities for each state's participation in the aforementioned absorption events are presented.

The meshes used in the FEM calculations of this study are not perfectly symmetrical, leading to a slight elimination of energy level degeneracy, which could be partially addressed by mesh refinement; however, considering the inherent asymmetry in nanostructures grown with the colloidal or other methods, further refinement is deemed unnecessary as it does not significantly enhance the physical accuracy of the models. This approach aligns with the empirical observation that real nanostructures, including those assumed to be spherical, exhibit subtle differences in excited energy levels, highlighting the practical limits of idealization in theoretical works.

We now turn our attention to the total intraband absorption exhibited by a nanoplatelet, as depicted in Figure 1. In the top left corner, the relationship between $\alpha(\omega,I)$ and $\hbar\omega$ is illustrated for a nanoplatelet structure, highlighting transitions between the ground state and the first three excited states. The transition from the ground state to the third excited state $|0\rangle \to |3\rangle$ is detailed in a separate inset to account for the varying energy ranges over which absorption occurs. Accompanying this, probability density plots for each state encircle the primary figure, providing a comprehensive view. Specifically, the ground state exhibits a single antinode with the electron predominantly localized at the nanoplatelet's center.

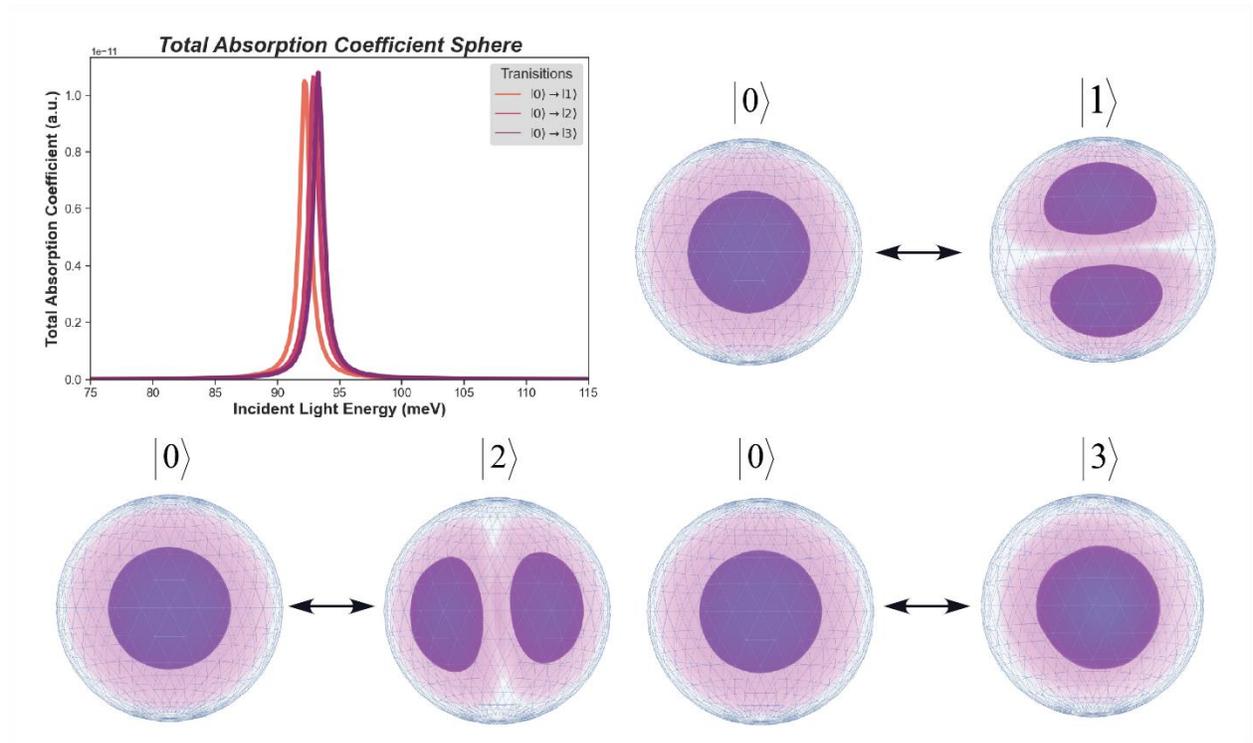

**Figure 2.** The dependence of the total intraband absorption coefficient on the incident light energy is presented in the left upper corner for the following transitions: $|0\rangle \to |1\rangle, |0\rangle \to |2\rangle, |0\rangle \to |3\rangle$ for an electron confined in an InAs spherical QD. In the plots surrounding the main figure, the probability densities for each state's participation in the aforementioned absorption events are presented.

The first and second excited states feature two antinodes, are diagonally oriented, and display quasi-degenerate characteristics, as evidenced by the positions and intensities of the $|0\rangle \to |1\rangle$ and $|0\rangle \to |2\rangle$ absorption curves. In contrast, the third excited state reveals four antinodes, indicating a significant increase in size quantization energy. This is attributed to the electron's enhanced interaction with the nanoplatelet walls. Notably, the total absorption curve for the

$|0\rangle \rightarrow |3\rangle$ transition demonstrates a blue shift from approximately $\sim 78 meV$, typical of the first two transitions, to around $\sim 155 meV$. This doubling in peak intensity is elucidated by the incorporation of $\hbar\omega$ into equations (1.2) and (1.3) linearly.

Proceeding to the examination of the well-researched yet fundamental case of spherical QDs, Figure 2 presents the intraband absorption curves for transitions from the ground state to the first three excited states ($|0\rangle \rightarrow |1\rangle, |0\rangle \rightarrow |2\rangle, |0\rangle \rightarrow |3\rangle$). These transitions exhibit predictable behavior, with the first three excited states being almost completely degenerate, displaying only minor discrepancies attributable to an anisotropic mesh. The absorption curves demonstrate a systematic blue shift corresponding to the ascending quantum order number, accompanied by an increase in peak intensity that mirrors this sequential order. Notably, the inherent spherical symmetry of the QDs results in maximal degeneracy, rendering the observed differences minute, on the order of fractions of a $meV$.

In stark contrast to the previously discussed nanoplatelet and spherical quantum dot structures, the nanorods present a unique case study in intraband absorption, primarily influenced by their aspect ratio which introduces novel quantum confinement effects. The absorption spectra for nanorods exhibit sharply defined peaks; particularly, the $|0\rangle \rightarrow |1\rangle$ transition displays a peak at approximately $24 meV$, indicative of a more confined electron in the ground state. This is further corroborated by probability density plots where the ground state shows a single elongated antinode, aligning with the rod's geometry, while the excited states, with two and three antinodes for the $|0\rangle \rightarrow |1\rangle$ and $|0\rangle \rightarrow |2\rangle$ transitions respectively, underscore the spatial confinement imposed by the nanorod's length. Notably, the $|0\rangle \rightarrow |3\rangle$ transition is marked by a significant blue shift, reflecting the heightened quantum confinement. Theoretical models are thus necessitated to factor in the nanorod's aspect ratio, as this modifies the confinement potential and consequently the energy levels for cylindrical quantum structures. The precise and predictable behavior of these absorption peaks relative to the aspect ratio not only advances our theoretical understanding but also opens avenues for experimentalists to exploit these properties in designing nanorods for specific optical applications, from filtering technologies to quantum computing components.

The absorption spectra of nanocones reveal intriguing characteristics that deviate from traditional quantum confinement behaviors observed in other nanostructures. Notably, the transition from the ground state to the first excited state $|0\rangle \rightarrow |1\rangle$ is red-shifted by approximately $50 meV$ in comparison to the transition to the second excited state $|0\rangle \rightarrow |2\rangle$, suggesting a lower confinement potential along the z-axis. This lower confinement correlates with the orientation of the first excited state, which extends primarily in the z-direction, thereby experiencing a reduced spatial restriction. On the other hand, the second and third excited states are confined closer to the base of the cone, leading to a pronounced size quantization effect. The confinement near the narrower base results in significantly higher energy states, as evidenced by the blue shift of the $|0\rangle \to |2\rangle$ and $|0\rangle \to |3\rangle$ transitions, with the latter being slightly blue-shifted relative to the former. This shift indicates an incremental increase in the quantization effect as the excited states ascend, aligning with the spatial gradient imposed by the conical geometry. These observations necessitate a refined theoretical framework that accounts for the directional dependence of quantum confinement in nanocones, providing new insights into electron localization and transition energies in tapered nanostructures. Such nuanced understanding is pivotal for tailoring the electronic and optical properties of nanocones for advanced nanotechnological applications, where directional control of electronic excitation is paramount.

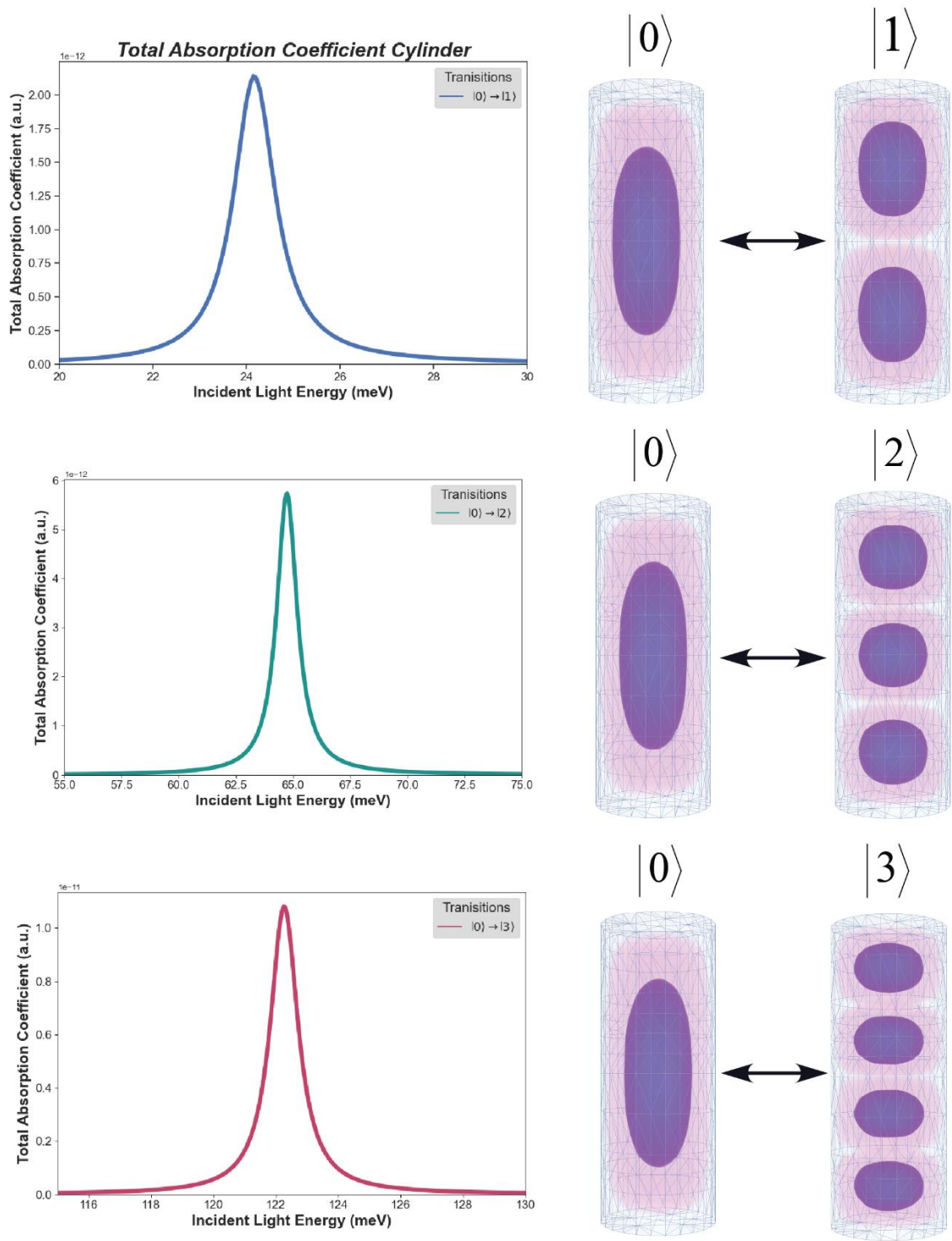

**Figure 3.** The dependences of the total intraband absorption coefficient on the incident light energy are presented on the left side of the figure besides the probability densities for each participant state for the following transitions $|0\rangle \rightarrow |1\rangle$, $|0\rangle \rightarrow |2\rangle$, $|0\rangle \rightarrow |3\rangle$ for an electron confined in an InAs nanorod.

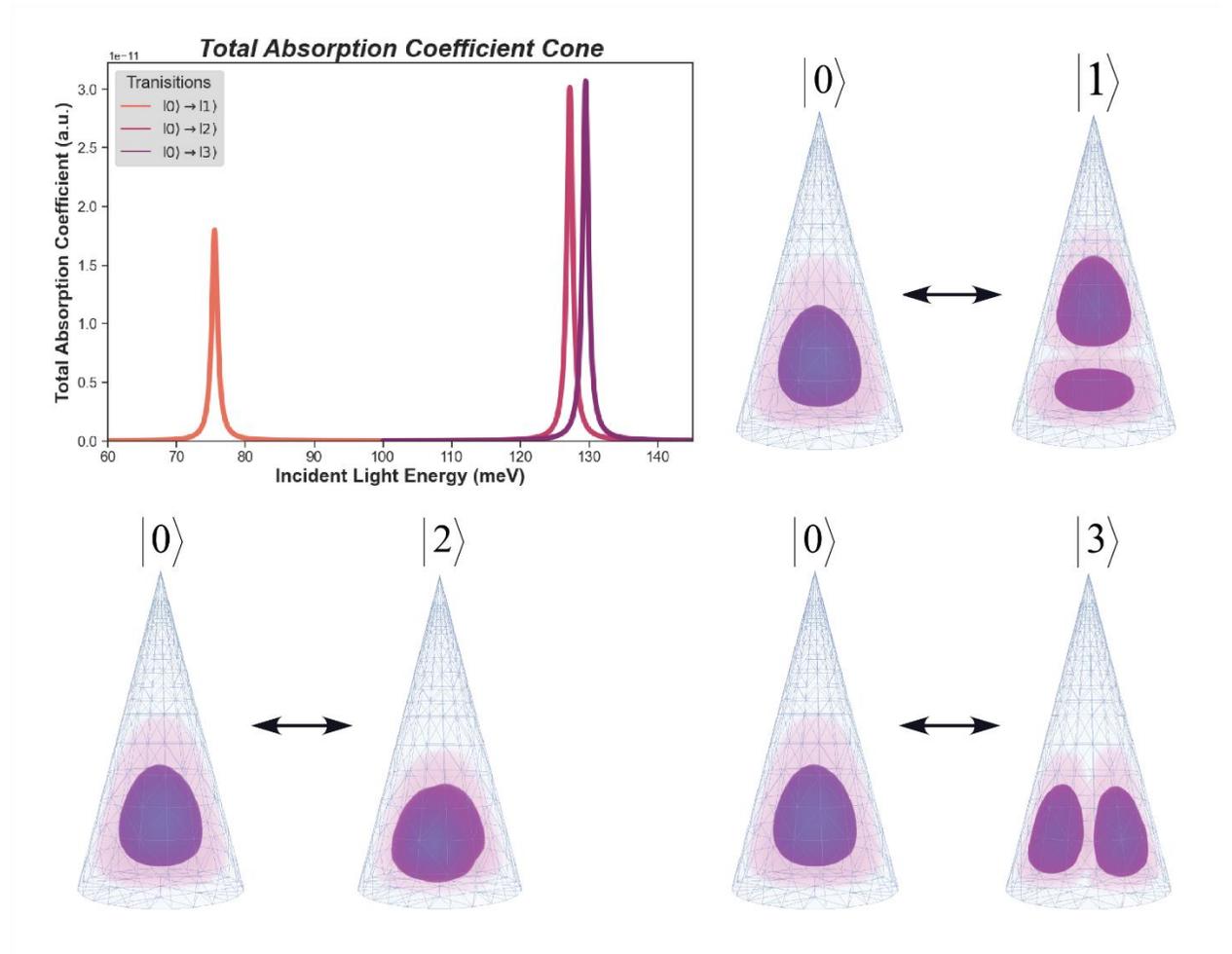

**Figure 4.** The dependence of the total intraband absorption coefficient on the incident light energy is presented in the left upper corner for the following transitions: $|0\rangle \to |1\rangle, |0\rangle \to |2\rangle, |0\rangle \to |3\rangle$ for an electron confined in an InAs nanocone. In the plots surrounding the main figure, the probability densities for each state's participation in the aforementioned absorption events are presented.

The absorption characteristics of nanotadpoles, as depicted in the accompanying figure, offer a compelling narrative that intertwines with the behaviors observed in nanocones. The results for nanotadpoles exhibit a similar quantum confinement effect, which is intricately linked to their geometry, akin to the trends identified in nanocones. The ground state to the first excited state transition $|0\rangle \to |1\rangle$ for the nanotadpole, marked by the initial peak, is red-shifted in a fashion that echoes the observations in nanocones, yet it appears at a slightly higher energy. This shift could be attributed to the bulbous base of the nanotadpole, which, while similar to the nanocone's base, offers a different spatial confinement due to the additional volume. The probability density of the first excited state suggests a localized electron presence within this base, as the state is oriented towards the larger cross-sectional area of the structure. The subsequent transitions to the second $|0\rangle \to |2\rangle$ and and third $|0\rangle \to |3\rangle$ excited states are characterized by a considerable blue shift, much like in the nanocone case.

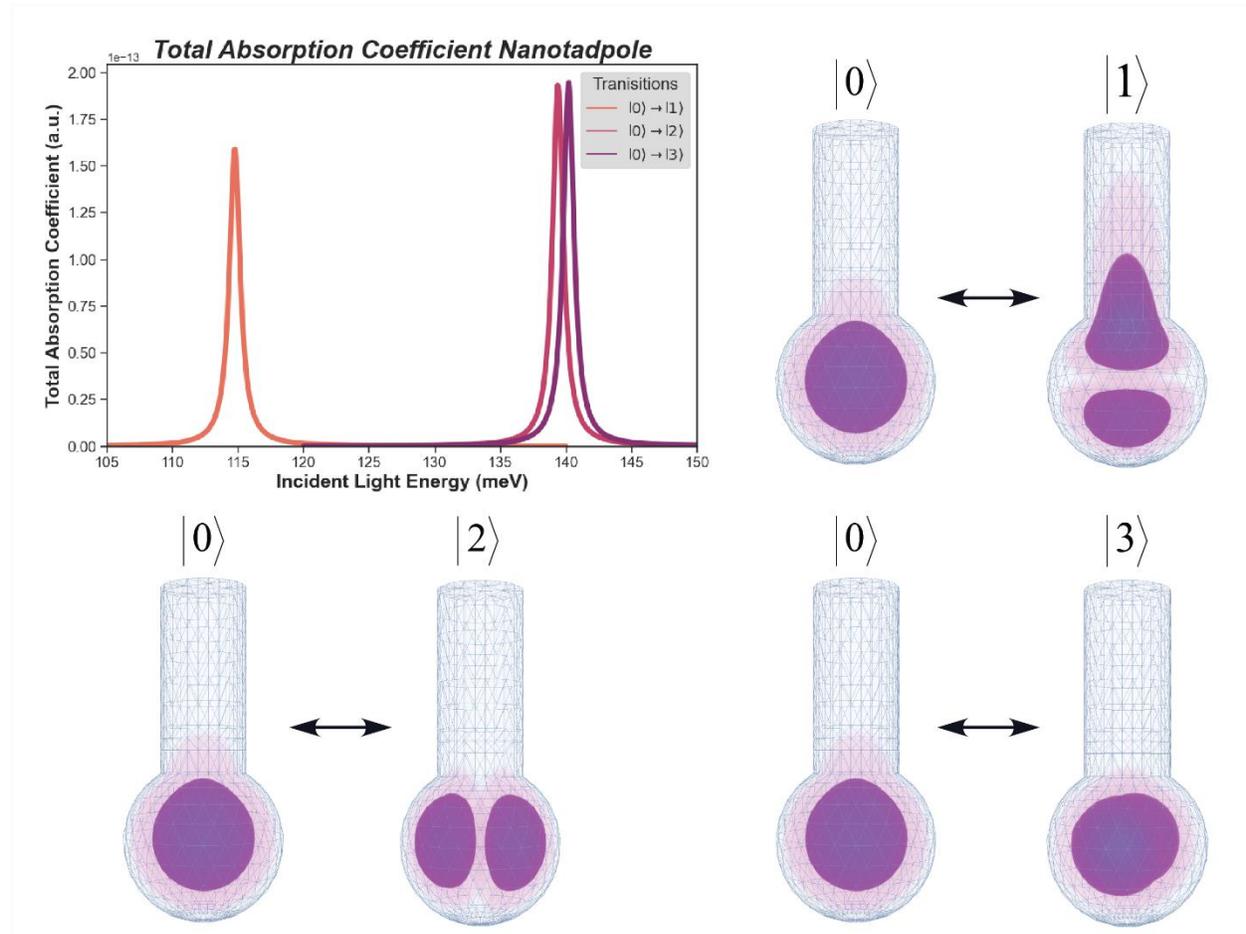

**Figure 5.** The dependence of the total intraband absorption coefficient on the incident light energy is presented in the left upper corner for the following transitions: $|0\rangle \to |1\rangle, |0\rangle \to |2\rangle, |0\rangle \to |3\rangle$ for an electron confined in an InAs nanotadpole. In the plots surrounding the main figure, the probability densities for each state's participation in the aforementioned absorption events are presented.

These shifts are more pronounced in the nanotadpole, indicating an even greater size quantization effect likely due to the elongated 'tail' portion of the structure. This part of the nanotadpole imposes a stronger confinement as the geometry tapers, akin to the nanocone, but with the added dimension of the 'tail' contributing to the electron's spatial confinement. Furthermore, the similarity in the spectral profiles between the nanocone and nanotadpole is striking, suggesting that the tail of the nanotadpole plays a significant role in the quantum confinement, similar to the effect of the tapering geometry in nanocones. However, the transitions in the nanotadpole's spectrum are sharper and more defined, which may be a result of the combined effects of the larger base and the elongated structure, contrasting with the uniform tapering of nanocones. In terms of theoretical implications, the nanotadpole's results support the notion that both the size and shape of the quantum structure critically influence the electron's energy states. This reinforces the need for theoretical models to adapt to complex geometries beyond simple spherical or cylindrical shapes, integrating the effects of both volume and geometric gradients.

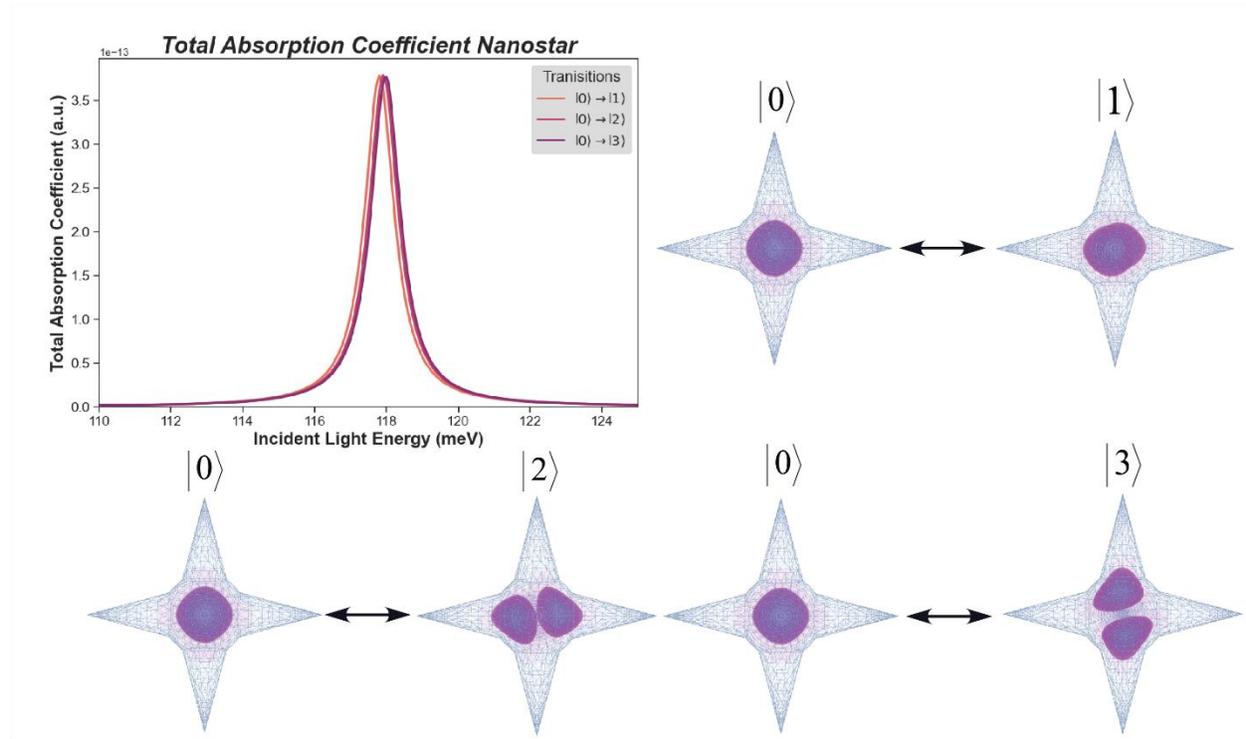

**Figure 6.** The dependence of the total intraband absorption coefficient on the incident light energy is presented in the left upper corner for the following transitions: $|0\rangle \rightarrow |1\rangle, |0\rangle \rightarrow |2\rangle, |0\rangle \rightarrow |3\rangle$ for an electron confined in an InAs nanostar. In the plots surrounding the main figure, the probability densities for each state's participation in the aforementioned absorption events are presented.

The nanostar, distinguished by its six symmetrical tips, stands out in its absorption characteristics when juxtaposed with other quantum structures. Figure 6 elucidates this by showcasing a series of intraband transitions $|0\rangle \rightarrow |1\rangle$, $|0\rangle \rightarrow |2\rangle$, $|0\rangle \rightarrow |3\rangle$ with closely aligned peak energies, a testament to the uniform confinement effects induced by the star's geometry. The diagonal orientation of the probability densities, distinct from the axial distributions observed in spherical QDs and nanotadpoles, suggests that the symmetry and spatial distribution of the nanostar's tips create an equilateral potential well that governs the electron behavior. This homogeneous confinement potential, unusual for such a complex structure, results in excited states that are remarkably similar in energy. The implication of this for quantum mechanics is profound, indicating that symmetry—even in a multifaceted shape like that of the nanostar—can lead to electronic states that mirror one another in energy levels, deviating from the conventional expectation that complexity begets energy diversity. This discovery not only advances our understanding of quantum confinement in nanostructures with multiple symmetrical protrusions but also hints at the potential for leveraging such geometries in developing nanoscale devices with consistent and predictable optoelectronic properties.

Analyzing the peak intensities of the $|0\rangle \rightarrow |3\rangle$ intraband transitions across a range of nanostructures within the context of total intraband absorption reveals a nuanced correlation between geometric form and optoelectronic response. Nanoplatelets, with their lower dimensionality, exhibit the most modest peak intensity at $10^{-18} a.u.$, underscoring the limitations imposed by their geometry on electron mobility and transition probability. This is in stark contrast to the three-dimensional spherical QDs and nanorods, both of which show significantly higher peak intensities of $10^{-11} a.u.$, suggesting a more favorable condition for electron transitions due to enhanced spatial confinement. The nanocone's peak intensity at $3 \times 10^{-11} a.u.$ surpasses that of its spherical and rod-shaped counterparts, likely reflecting an interplay between its tapering geometry and an increased overlap of electron wavefunctions, which might also amplify the nonlinear absorption due to stronger local electric fields. Conversely, the intricately shaped nanotadpole and nanostar, with peak intensities at $2 \times 10^{-13} a.u.$ and $3.5 \times 10^{-13} a.u.$ respectively, hint at a more complex dynamic where elaborate structures could constrain electron overlap and diminish linear absorption, yet potentially enhance nonlinear interactions with incident light. Collectively, these values not only inform the design and potential applications of nanostructures in optoelectronic devices but also illustrate the intricate interplay between structure and electronic properties in nanoscale materials.

## 4. Conclusion

In this study, we explored the dynamics of total intraband absorption in various nanostructures—nanoplatelets, spherical quantum dots, nanorods, nanocones, and nanostars—highlighting the significant influence of geometric configuration on quantum confinement effects and optoelectronic properties. Nanoplatelets underscored the foundational role of dimensionality, while spherical QDs and nanorods illustrated how geometric symmetry and aspect ratios impact excited state energy levels. The investigation into nanocones and nanostars revealed unexpected electronic behaviors due to their complex geometries, such as tapering and symmetrical protrusions, respectively. This contrast in absorption characteristics across different structures emphasizes the nuanced interplay between structure and electronic properties, contributing to our understanding of quantum mechanics in complex nanostructures and paving the way for the development of tailored nanomaterials for optoelectronic applications. Our findings underscore the importance of a refined theoretical framework that considers both geometric and electronic complexities, offering new avenues for innovation in nanoscale engineering and promising exciting advancements in nanotechnology and its applications.

## 5. Acknowledgement

This work was supported by the **RA MESCS Higher Education and Science Committee (Research project № 23RL-1B004)** and Horizon—2020 research and innovation program of the European Union (grant no. 952335, NanoQIQO project).